\documentclass{article}
\pdfoutput=1

\usepackage{arxiv}

\usepackage[utf8]{inputenc} 
\usepackage[T1]{fontenc}    
\usepackage{hyperref}       
\usepackage{url}            
\usepackage{booktabs}       
\usepackage{amsfonts}       
\usepackage{nicefrac}       
\usepackage{microtype}      
\usepackage{graphicx}
\usepackage{gensymb}
\usepackage{xcolor}
\usepackage{mathtools}
\usepackage{upgreek}
\usepackage{textcomp}
\usepackage{siunitx}

\title{Sampling the Cu-Fe-Co phase diagram by severe plastic deformation for enhanced soft magnetic properties}

\author{Martin St\"uckler$^{a,1}$, Stefan Wurster$^a$, Lukas Weissitsch$^a$, Heinz Krenn$^b$, Reinhard Pippan$^a$, Andrea Bachmaier$^a$ \\
$^a$   Erich Schmid Institute of Materials Science, Austrian Academy of Sciences,  Jahnstra{\ss}e 12, 8700 Leoben, Austria \\
$^b$    Institute of Physics, University of Graz, Universit\"atsplatz 5, 8010 Graz, Austria\\
$^1$   \texttt{martin.stueckler@oeaw.ac.at}
  }

\begin{document}
\maketitle
\begin{abstract}
Initial powder mixtures of Cu, Fe and Co are exposed to severe plastic deformation by high-pressure torsion to prepare solid solutions. A broad range of compositions is investigated, whereas this study aims at the synthesis of soft magnetic materials and therefore at the formation of a homogeneous and nanocrystalline microstructure. For intermediate ferromagnetic contents, high-pressure torsion at room temperature yields single-phase supersaturated solid solutions. For higher ferromagnetic contents, two consecutive steps of high-pressure torsion deformation at different temperatures yield the desired nanocrystalline microstructure. Depending on the Co-to-Fe-ratio, either a single-phase supersaturated solid solution or a nanocomposite forms. The composite exhibits an enhanced magnetic moment, indicating the formation of a (Fe,Co)-alloy upon severe plastic deformation. Soft magnetic properties are verified for large Co-to-Fe-ratios and this microstructure is found to remain stable up to 400 \degree C.
\end{abstract}


\section{Introduction}
Nanocrystallinity is a prerequisite to form high performance soft magnetic materials \cite{herzer2013modern}, whereas the enhanced number of interfaces gives rise to diffusion-mediated magnetic anisotropy making an easy tunability of the magnetic hysteresis loop possible \cite{herth2004interface, herth2004diffusion}. This large number of diffusion paths \cite{gleiter1989nanocrystalline} has attracted interest in the synthesis of energy materials where enhanced diffusion gives rise to improved materials properties for various applications such as carbon dioxide adsorption \cite{baiano2020role}, synthesis of molecular hydrogen \cite{udayabhanu2020one} or energy storage applications \cite{pedico2018high}. \newline
With high-pressure torsion (HPT), a technique of severe plastic deformation (SPD), nanocrystalline materials can be processed in bulk form \cite{valiev2000bulk}. Furthermore, bulk state reactions can be induced by HPT, leading to metastable states which can even be retained after pressure release \cite{kilmametov2018alpha, bachmaier2019high, han2019bulk}. Regarding the production of magnetic materials, it has been shown that HPT is capable to tune the coercivity even by orders in magnitude with respect to the initial state, whereas larger effects have been observed for materials consisting of more than one chemical element \cite{straumal2009effect, scheriau2010magnetic, edalati2013high, bachmaier2017tailoring, bachmaier2019high}. To achieve low coercivities, the formation of a homogeneous microstructure, exhibiting nanocrystalline grains with low magnetocrystalline anisotropy, is desirable.\newline
In Ref.~\cite{stuckler2019magnetic}, it has been shown that desired microstructures can be processed from binary powder mixtures of Cu-Fe and Cu-Co, whereas Cu-Co samples exhibit soft magnetic properties. The processing limits of HPT at room temperature have been examined as well: in case of Cu-Fe, it was shown that homogeneous deformation was achieved for Fe contents below 25~wt.\%. Higher Fe-contents yield localization of strain resulting in an inhomogeneous, multiphase microstructure. In case of Co-Cu, supersaturated solid solutions have been processed for Co contents between 28~wt.\% and 67~wt.\%. Lower Co-contents yield comparatively large, residual Co particles, impeding improved magnetic properties. Higher Co contents result in brittle specimens with cracks forming during deformation. \newline
\begin{figure*}[t]
\begin{center}
\includegraphics[width=0.9\linewidth]{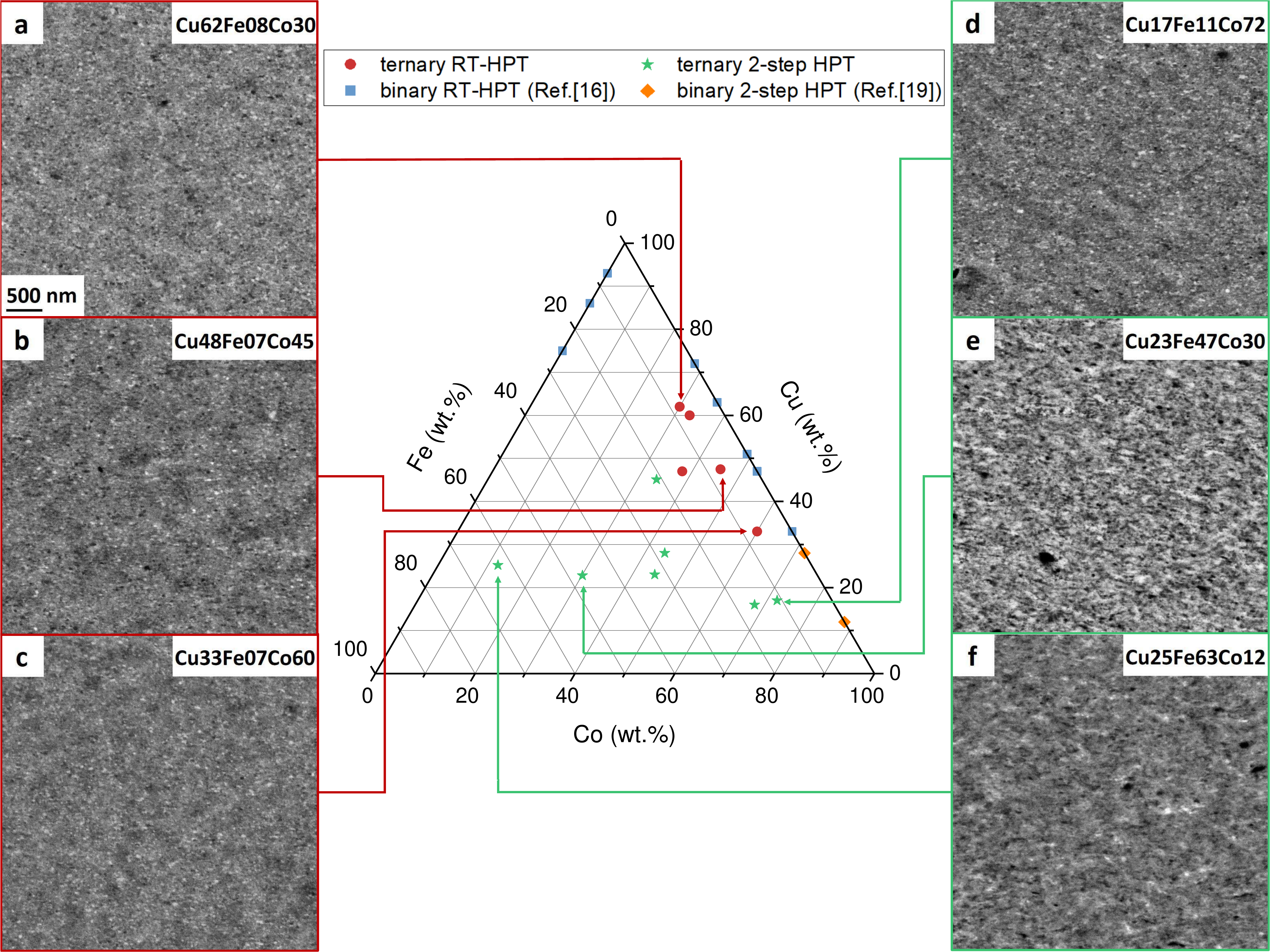}
\end{center}
\caption{SEM micrographs of as-deformed samples. The ternary phase diagram shows the samples investigated in this study with the compositions determined from EDS-measurements. Images on the left (a)-(c) show samples, which are processed by RT-HPT. Images on the right (d)-(f) show samples processed by 2-steps of HPT as described in the text. Micrographs were taken on HPT-discs cut in half at $r$~=~3~mm with the radial direction parallel to the horizontal axis. The scale bar in (a) applies to all micrographs.}
\label{fig:SEM_asdef}
\end{figure*}
This study aims at the production of nanocrystalline, bulk materials of the ternary Cu-Fe-Co system, which is expected to show superior soft magnetic properties in comparison to the Cu-Co system. In the present work, different Fe-Co ratios are mixed with Cu, since the addition of Fe to Co leads for certain compositions to a decreasing magnetocrystalline anisotropy and increasing saturation magnetization \cite{LandoltBornstein1986, kuhrt1992formation}. \newline
The focus of the study is on the processability of Cu-Fe-Co alloys by HPT and the formation of homogeneous microstructures, also involving the formation of supersaturated states, by tailored processing routes. Furthermore, the soft magnetic properties of the most promising ternary materials are investigated. For achieving the goal of homogeneously deformed, nanocrystalline ternary Cu-Fe-Co materials of high ferromagnetic content, room temperature HPT (RT-HPT) is not the method of choice, due to the above-mentioned problems encountered in the binary systems. A two-step process involving deformation at two different temperatures, as it is proposed in \cite{stuckler2020mfm} for Cu-Co of very high Co-content, leads to better results. \newline
Furthermore, it is well known that nanocrystalline materials are prone to microstructural changes, which happens already at very low homologous temperatures. As this has a strong influence on the magnetic properties, the thermal stability of the as-deformed microstructures is investigated.
\section{Experimental}
Samples were processed from conventional powders (Fe: MaTeck 99.9\% -100+200 mesh; Co: GoodFellow 99.9\% 50-150 $\mu$m; Cu: Alfa Aesar -170+400 mesh 99.9\%). Powder mixtures were consolidated under hydrostatic pressure in Ar-atmosphere. The consolidated powders were exposed to SPD by HPT, resulting in cylindrical specimens (diameter: 8~mm; thickness: 1~mm). For deformation at higher temperatures (up to 500 \degree C), the anvils were inductively heated. HPT was performed under a pressure of 5~GPa at 1.28~min$^{-1}$ rotational speed of the anvils. To ensure a microstructural steady state, 50-150 numbers of turns were applied in total, resulting in a maximum achieved shear strain $\gamma$ of 1000-3000 at a radius $r$ of 3~mm. During room temperature (RT) deformation the sample is cooled by compressed air avoiding heating. Microstructural investigations were performed by scanning electron microscopy (SEM; Zeiss LEO1525) in backscattered electron (BSE) mode. SEM micrographs are taken at equal settings throughout the whole study (acceleration voltage, BSE-gain, aperture size, current mode). The local compositions of the samples were determined by an energy dispersive X-ray spectroscopy system (EDS; Bruker XFlash 6\textbar 60) by averaging over a set of at least 30 individual spots and is given in weight percent (wt.\%; $\pm$1~wt.\%) herein. Quantitative grain size evaluation is carried out by Transmission Kikuchi Diffraction (TKD) using a Bruker e$^-$-Flash$^{FS}$ electron backscatter diffraction detector. Vickers hardness measurements were performed with a Buehler Micromet 5100 hardness testing device under a load of 0.5~kg (HV0.5). The crystallographic structure was measured by X-ray diffraction (XRD) using a Bruker D2 Phaser (Co-K$_\alpha$ radiation) and by synchrotron-XRD (DESY; PETRA III: P07) in transmission mode at an energy of 100~keV. DC-magnetometric measurements were performed with a Quantum Design MPMS-XL7 SQUID-magnetometer at 8~K.
\begin{figure*}
\begin{center}
\begin{tabular}{cc}
\includegraphics[width=0.5\linewidth]{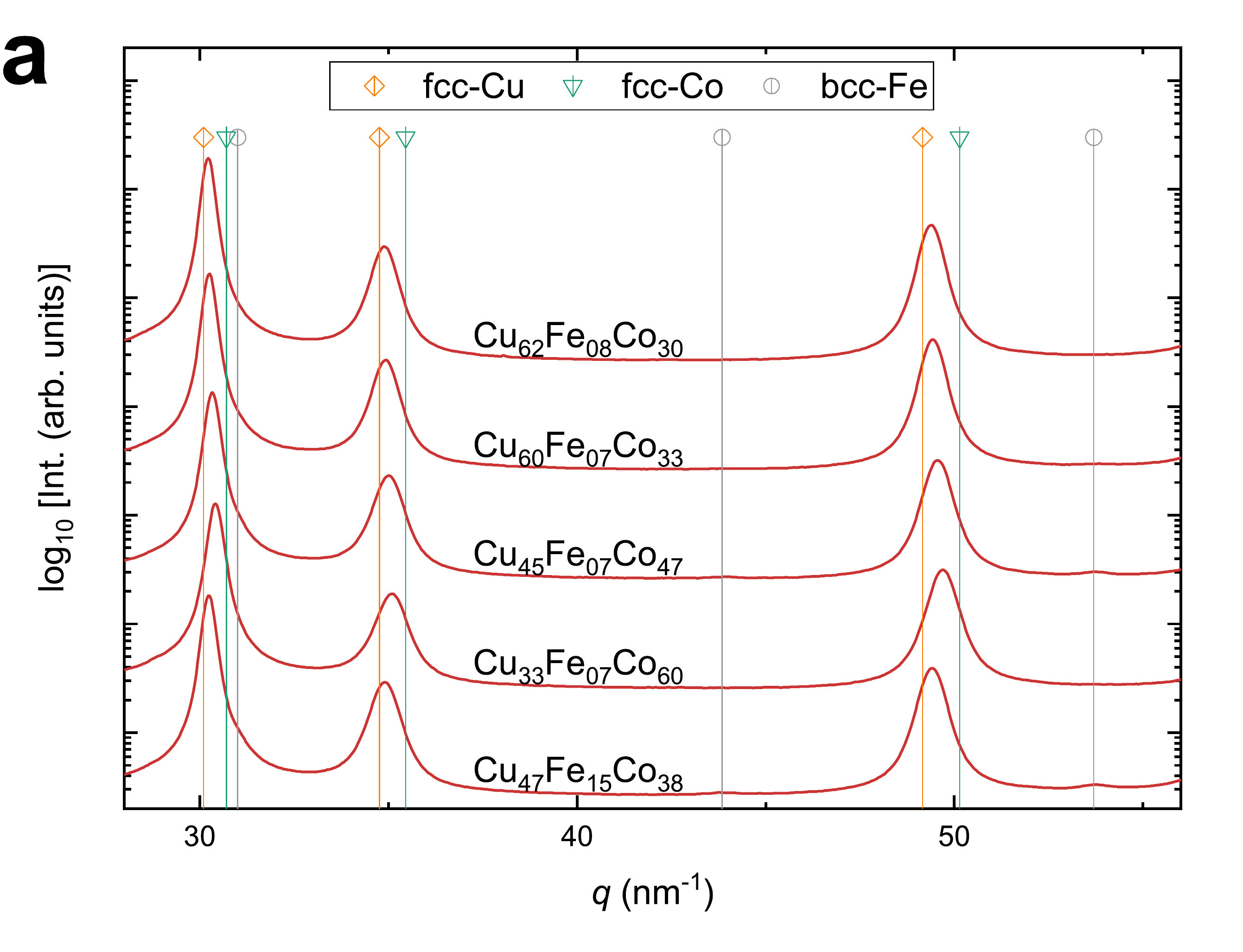} &
\includegraphics[width=0.5\linewidth]{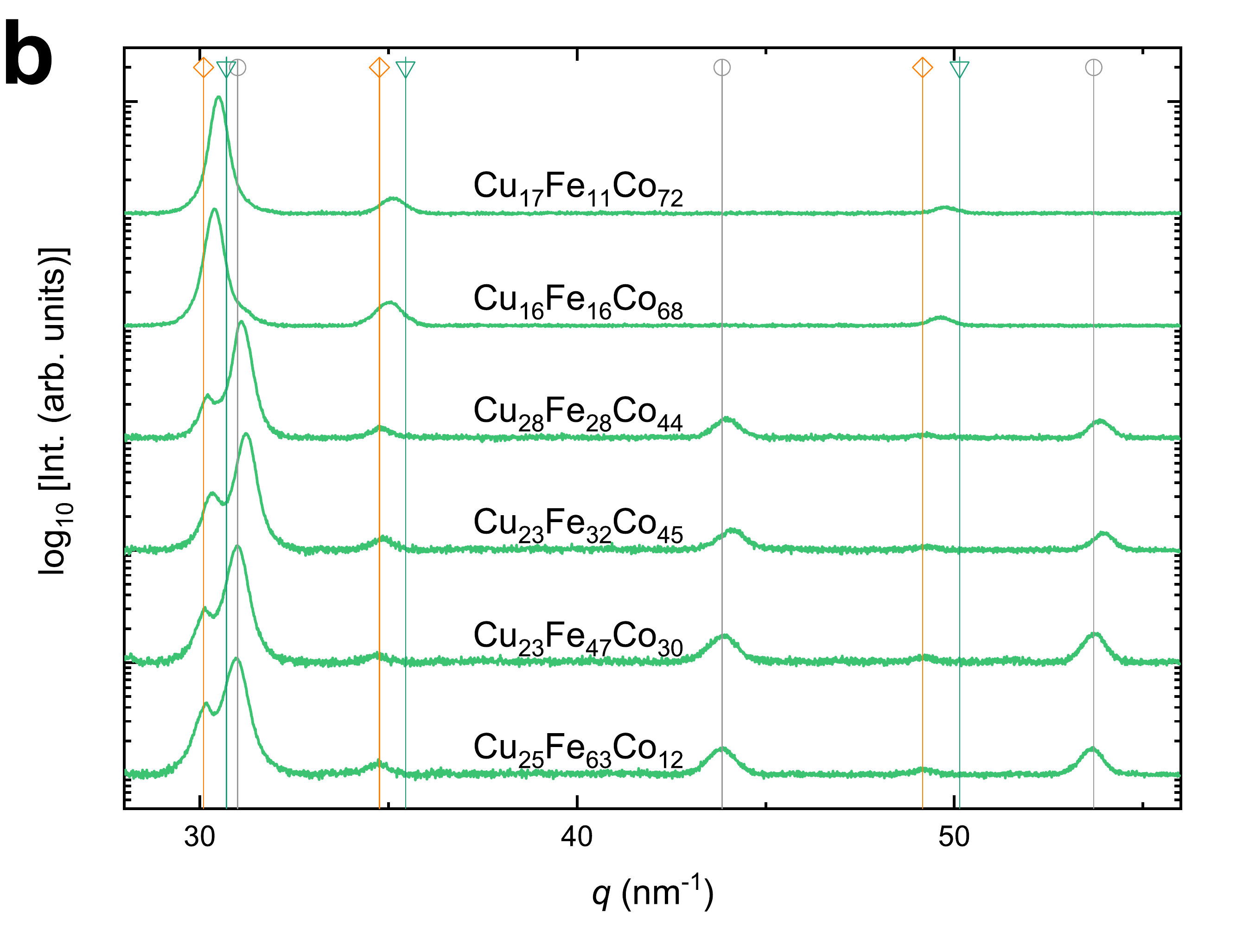}
\end{tabular}
\end{center}
\caption{XRD patterns of as-deformed samples (see Fig.~\ref{fig:SEM_asdef}). In (a), synchrotron XRD patterns are shown for RT-HPT processed samples. In (b), XRD patterns for 2-step HPT-processed samples are shown (see Fig.~\ref{fig:SEM_asdef}), measured with Co-K$_\alpha$ radiation. An increasing Fe/Co-ratio is shown from top to bottom (b). The legend in (a) applies also to (b).}
\label{fig:xrd}
\end{figure*}
\section{Results and Discussion}
\subsection{Synthesis of soft magnetic ternary materials by HPT}
The ternary phase diagram (Fig.~\ref{fig:SEM_asdef}) shows the compositions as determined by EDS investigated in this study. Compositions which have been processed by HPT-deformation at RT (50 turns) are marked by red dots and are referred to as RT-HPT in the following. The corresponding SEM micrographs are shown in Fig.~\ref{fig:SEM_asdef}(a)-(c) and reveal feature sizes in the nanocrystalline regime. Furthermore, low phase contrast is apparent, indicating a chemical homogeneity, which is also confirmed by XRD-measurements (see Fig.~\ref{fig:xrd}(a)). The RT-HPT deformed samples exhibit compositions within the limits of RT-processability of the respective binary systems \cite{stuckler2019magnetic}. Initial studies on the processability of ternary mixtures by RT-HPT showed the same trend as was found for binary mixtures. Indeed, samples exhibiting Fe contents higher than 20~wt.\% and/or Co contents above 70~wt.\% show residual particles and the formation of shear band or cracks. Thus, one is restricted in the useable ternary compositions for RT-HPT, but especially high Fe- and Co-containing materials are strived for processing soft magnetic materials. Consequently, to overcome this problem, the same technique as it was used for binary Cu-Co with very high Co-contents \cite{stuckler2020mfm} will be applied.
These samples are referred to as 2-step HPT samples in the following. At first, the deformation is performed at elevated temperatures (300~\degree C or 500~\degree C; 100 turns or 50 turns), achieving a homogeneously deformed microstructure for samples with high ferromagnetic content. After, the sample is exposed to a second deformation step at RT (additional 50 turns), refining the already homogeneous microstructure. The 2-step HPT samples are indicated by green stars in Fig.~\ref{fig:SEM_asdef}. Apparent phase contrast is visible in the SEM micrographs in Fig.~\ref{fig:SEM_asdef}(e) and (f), indicating chemically different phases. Furthermore, a larger feature size can be noticed in comparison to the specimen exhibiting the lowest Fe-content (Fig.~\ref{fig:SEM_asdef}(d)).\newline
Fig.~\ref{fig:xrd}(a) shows results from synchrotron-XRD experiments for RT-HPT processed samples. All samples exhibit a major fcc pattern. Beside a very shallow bcc-Fe peak for Cu$_{47}$Fe$_{15}$Co$_{38}$ at $q$~=~53.6~nm$^{-1}$, only single-phase crystallographic structures are present, as it has been reported for binary Cu-Co \cite{stuckler2019magnetic}. In Fig.~\ref{fig:xrd}(b), XRD patterns of the 2-step HPT-deformed samples are shown. For an increasing Fe/Co-ratio (from top to bottom), a transition from single phase to dual phase crystallographic structures can be observed in accordance to SEM micrographs in Fig.~\ref{fig:SEM_asdef}(e),(f). This is consistent with the binary Co-Fe phase diagram, showing a transition from fcc to bcc for increasing Fe-contents \cite{ohnuma2002phase, gilles2010probing}. For Co-contents $\geq$ 60~wt.\%, the fcc-peaks shift from the Cu-peak position, indicating the formation of supersaturated solid solutions. In contrast, for lower Co-contents this peak shift is hardly observed, indicating the presence of a rather pure fcc-Cu phase. We assume that $\alpha$-(Fe,Co) forms as a second phase, according to the thermodynamical equilibrium phase diagram. However, the results demonstrate that HPT is capable of synthesizing single-phase supersaturated solid solutions well beyond the solubility limits given at thermodynamical equilibrium \cite{bamberger2002evaluation}.\newline 
In all XRD-patterns, a huge peak width can be observed, arising from the large amount of crystal defects, but also from grain refinement \cite{Ungar1999, Ungar2007using}. To quantify the grain size, a RT-HPT processed sample (Cu$_{60}$Fe$_{07}$Co$_{33}$) was measured with TKD, serving as an example for the grain sizes obtained in this study. Five scans were jointly analyzed, revealing a median grain size of 66~nm, which is a somewhat smaller grain size as has been obtained for binary Cu-Co (76~nm-100~nm \cite{stuckler2019magnetic}).\newline
\begin{figure}
\begin{center}
\includegraphics[width=0.7\linewidth]{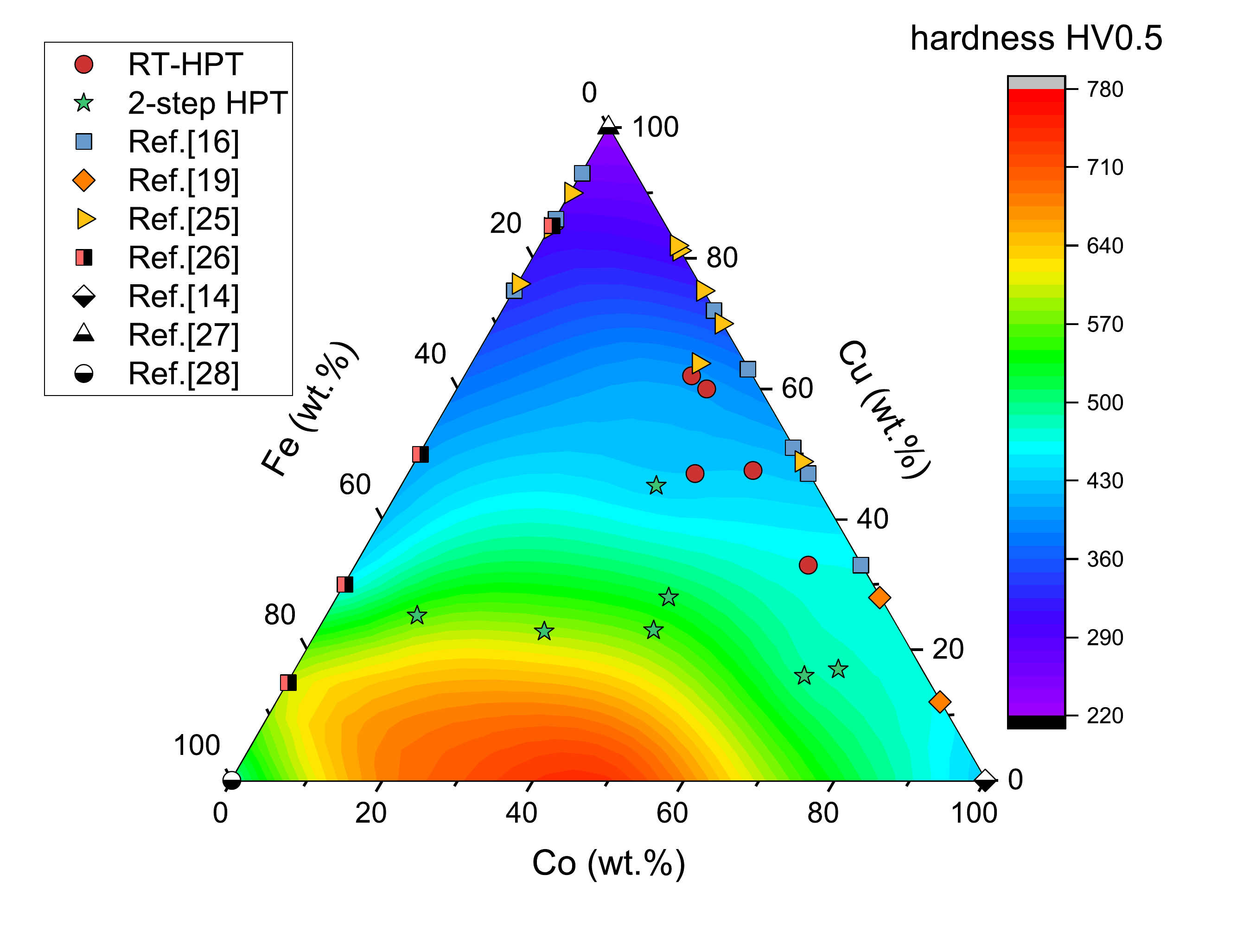}
\end{center}
\caption{Steady state Vickers hardness as a function of composition for HPT-processed mixtures of Cu-Fe-Co. For the samples investigated in this study, the mean hardness for indents between 2~mm $\leq$ $r$ $\leq$ 3.5~mm is plotted.}
\label{fig:hardness}
\end{figure}
Vickers hardness is very sensitive to changes in the microstructure and is used to probe the microstructural homogeneity of the HPT-deformed materials. Therefore, hardness testing is performed as a function of radius for each sample. For all samples investigated in this study the applied deformation ensures a microstructural saturation for $r$~$\geq$~2~mm. In Fig.~\ref{fig:hardness}, the mean hardness for indents between 2~mm $\leq$ $r$ $\leq$ 3.5~mm is plotted as a function of composition. To complete the hardness plot for this ternary system Fe-Co specimens without any Cu were processed using the 2-step HPT process. The data from this study are compared with hardness values for binary HPT-deformed samples from other studies. In all these studies, initial powder mixtures are used for multi-phase materials \cite{stuckler2019magnetic, stuckler2020mfm, wurster2019tuneable, bachmaier2012}. Pure elements have been HPT-deformed from powders \cite{edalati2013high, kormout2017CuAg} or from bulk ARMCO Fe \cite{valiev1996}. All mentioned studies have in common that at least the last processing step has been carried out at RT, yielding the RT saturation grain size for every composition \cite{renk2019saturation}. Reaching the saturation grain size is accompanied by a steady state hardness as well \cite{kormout2017deformation, stuckler2020intermixing}. The grain refinement upon HPT-deformation causes an increase in hardness according to Hall-Petch hardening \cite{hall1951deformation, petch1953cleavage}. 
A monotonic increasing hardness can be observed for Cu-Co up to about 80~wt.\% of Co. Larger Co-contents exhibit smaller hardness values which is expected to arise from a diminishing contribution of solid solution hardening. A similar trend is visible for binary Cu-Fe. In comparison to binary Cu-Co, the hardness is increased in the ternary alloys. As it has been shown in the XRD-patterns (see Fig.~\ref{fig:xrd}), two crystallographic phases start to form, when exceeding a certain Fe-content. For dual phase materials, a rapid increase in hardness can be observed, arising from a complex hardening behavior involving phase and grain boundaries \cite{WERNER1985175}.
\subsection{Magnetic properties of ternary materials}
\begin{figure}
\begin{center}
\includegraphics[width=0.5\linewidth]{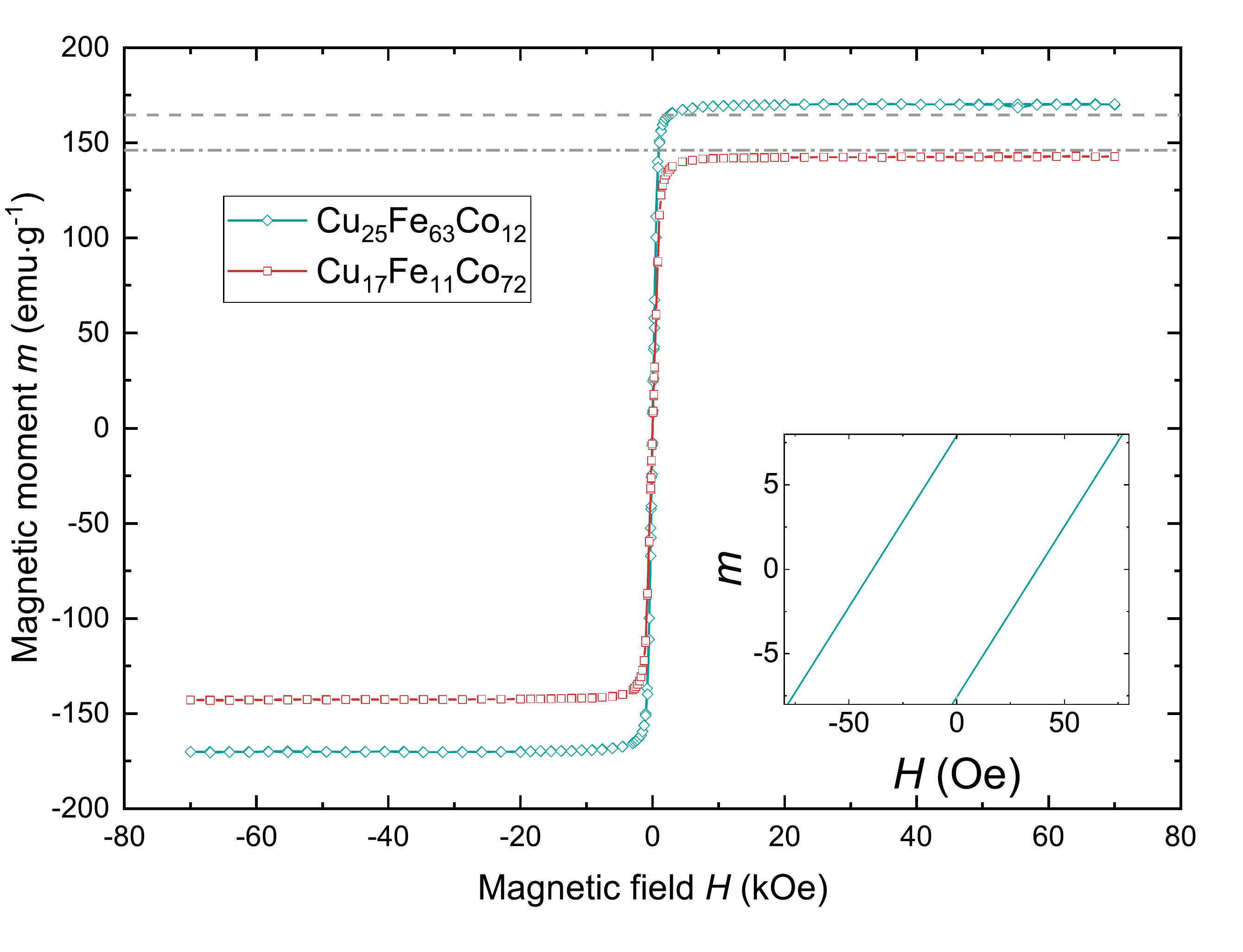}
\end{center}
\caption{Hysteresis curves measured by SQUID-magnetometry at 8~K of two samples, processed by 2-step HPT deformation process. The dashed lines give literature values for isolated Fe and Co phases for the compositions investigated. The inset shows Cu$_{25}$Fe$_{63}$Co$_{12}$ at small fields, whereas for Cu$_{17}$Fe$_{11}$Co$_{72}$ the coercivity collapses to such a small value, which cannot be resolved due to residual fields in the superconducting magnet.}
\label{fig:msat}
\end{figure}
The magnetic moments $m$ of one Fe-rich and one Co-rich specimen, containing about 20~wt\% of Cu, namely Cu$_{25}$Fe$_{63}$Co$_{12}$ and Cu$_{17}$Fe$_{11}$Co$_{72}$, are measured by SQUID-magnetometry at 8~K. As it has been shown, whilst the Fe-rich material represents a dual phase material, the Co-rich material is composed of a single phase. Since intermetallic alloys of (Fe,Co) exhibit an overproportional high magnetic moment \cite{kuhrt1992formation}, the alloying behavior on the atomic scale can be investigated by magnetic measurements. In Fig.~\ref{fig:msat}, the magnetic moments $m$ are plotted as a function of applied field $H$ for both compositions. For the Fe-rich composition Cu$_{25}$Fe$_{63}$Co$_{12}$, $m$ is larger than the reference value for isolated Fe and Co, showing the formation of a (Fe,Co)-phase in accordance with XRD-data (see Fig.~\ref{fig:xrd}). For the Co-rich composition Cu$_{17}$Fe$_{11}$Co$_{72}$, the saturation magnetization coincides with a linear superposition of reference values for Fe and Co (plotted as horizontal lines in Fig.~\ref{fig:msat}). Therefore, no formation of a (Fe,Co)-phase is apparent in this case, matching with the microstructural investigations, which show the presence of a single-phase fcc microstructure (see Fig.~\ref{fig:SEM_asdef},\ref{fig:xrd}).\newline
The Fe-rich composition Cu$_{25}$Fe$_{63}$Co$_{12}$ exhibits a large coercivity of $H_C$~=~(39$\pm$1)~Oe, whereas the coercivity of Co-rich Cu$_{17}$Fe$_{11}$Co$_{72}$ is smaller than the residual field of the superconducting magnet and cannot be properly resolved \cite{ney2018tutorial}, but is expected to be smaller than 10~Oe. An exhaustive determination of the coercivity for Cu$_{17}$Fe$_{11}$Co$_{72}$ is instead performed by means of AC-hysteresis measurements and can be found in Ref.~\cite{stueckler2021insitu}, revealing a $H_C$ of about 2~Oe at RT. The high $H_C$ for Cu$_{25}$Fe$_{63}$Co$_{12}$ is due to the larger grain size, as well as the comparatively larger magnetocrystalline anisotropy for this intermetallic alloy of (Fe,Co) \cite{LandoltBornstein1986}.\newline
According to the random anisotropy model, a breakdown in coercivity is expected as the grain size $d$ falls below a certain limit \cite{herzer2013modern}. However, the low coercivity for the Co-rich composition Cu$_{17}$Fe$_{11}$Co$_{72}$ cannot be explained by grain refinement only, since similar feature sizes have been observed in all samples investigated. The transition from the classical scaling behavior to random anisotropy is controlled by the magnetocrystalline anisotropy $K$ and the magnetoelastic stiffness constant $A$ [eq.~\ref{eq:random_ani}].
\begin{equation}
  H_C \propto
    \begin{cases}
      1/d & \text{for} \: d > \sqrt{A/K}\\
      d^6 & \text{for} \: d < \sqrt{A/K}
    \end{cases}
\label{eq:random_ani}
\end{equation}
Solid solutions of (Fe,Co) exhibit smaller magnetocrystalline anisotropies compared to pure Co \cite{LandoltBornstein1986}, causing a transition to random anisotropy at larger $d$. Furthermore, in a previous study on the magnetic properties of HPT-deformed Co$_{72}$Cu$_{28}$ \cite{stuckler2020mfm} it has been discussed that the soft magnetic properties might originate from an alloying effect, i.e. the formation of a supersaturated solid solution. The even smaller $H_C$ for the Co-rich Cu$_{17}$Fe$_{11}$Co$_{72}$ \cite{stueckler2021insitu} is therefore expected to originate from a lowering in the magnetocrystalline anisotropy due to the addition of Fe with respect to the binary composition and might be further lowered by the formation of a supersaturated solid solution including Cu. Similarily, a  significant reduction in coercivity has also been observed for single-phase solid solutions of Co-Fe-Ni synthesized by mechanical alloying \cite{betancourt2020structural}. The rather large $H_C$ in the Fe-rich Cu$_{25}$Fe$_{63}$Co$_{12}$ indicates that for this composition the transition to random anisotropy is not yet reached.\newline
The results show that, although being similarly processed, the magnetic behavior of both compositions differs extremely, arising from different microstructural states after HPT-deformation.
\subsection{Evolution of the supersaturated state upon annealing}  
\begin{figure}
\begin{center}
\begin{tabular}{cc}
\includegraphics[width=0.24\linewidth]{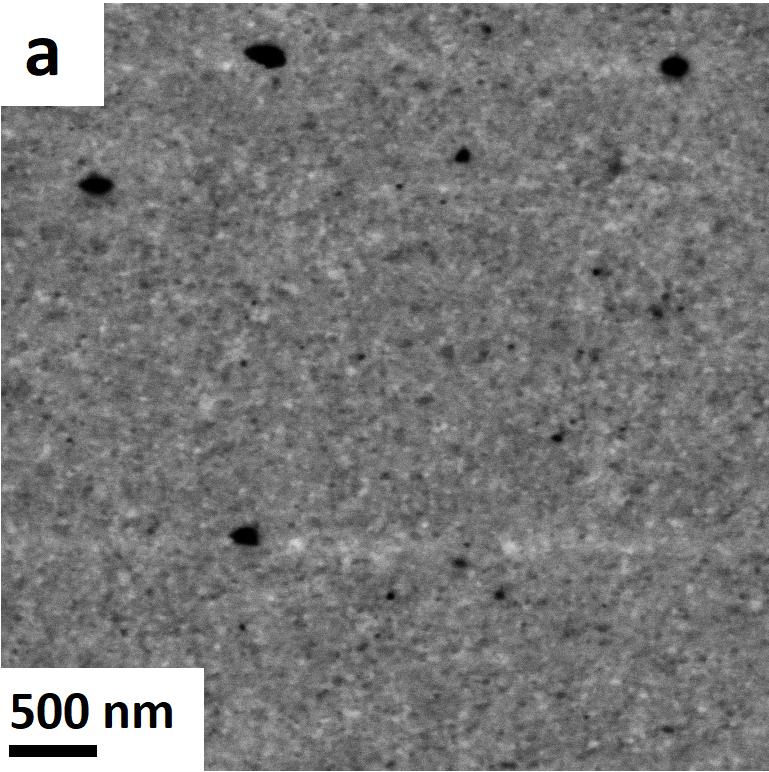} &
\includegraphics[width=0.24\linewidth]{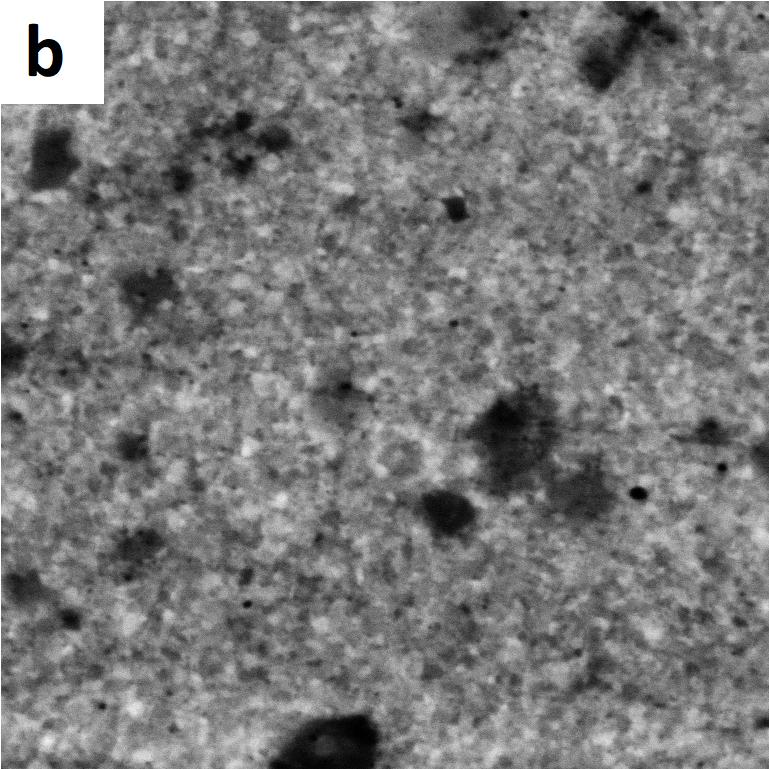}\\
\includegraphics[width=0.24\linewidth]{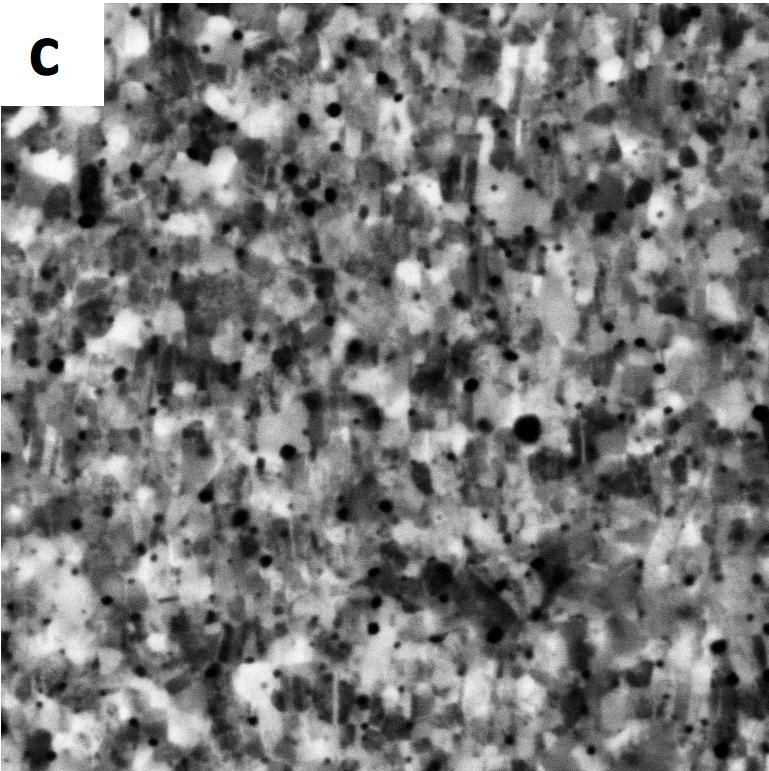} &
\includegraphics[width=0.24\linewidth]{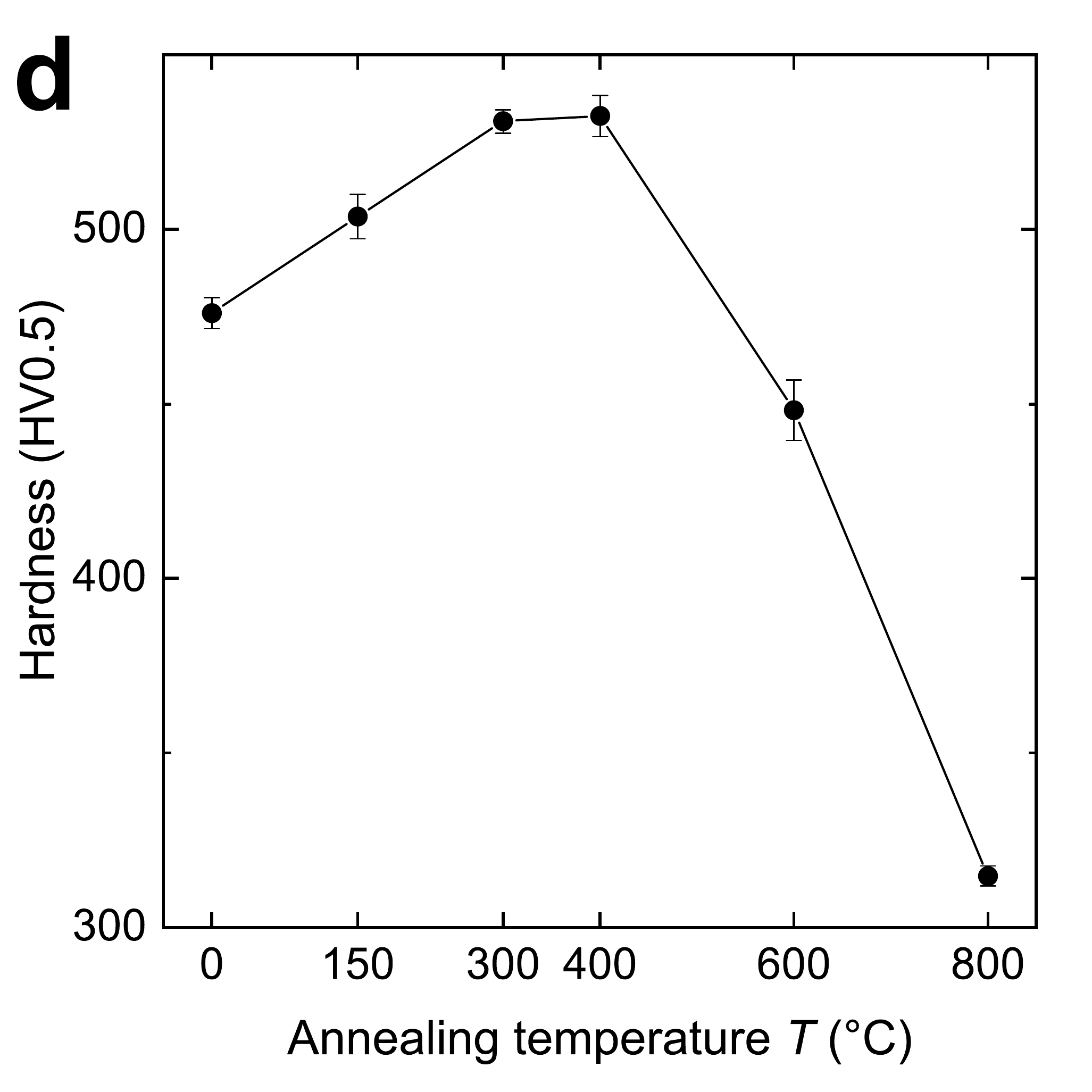}
\end{tabular}
\end{center}
\caption{SEM micrographs of Cu$_{17}$Fe$_{11}$Co$_{72}$ (see Fig.~\ref{fig:SEM_asdef}(d)) exposed to annealing temperatures at 300~\degree C (a), 600~\degree C (b) and 800~\degree C (c). The scale bar in (a) applies also to (b) and (c). In (d), Vickers hardness is plotted as a function of annealing temperature.}
\label{fig:SEM2}
\end{figure}

Since supersaturated solid solutions are prone to microstructural changes occurring already at low temperatures, one composition is investigated regarding its thermal stability. The 2-step HPT-processed sample of lowest coercivity and with composition Cu$_{17}$Fe$_{11}$Co$_{72}$ forms a single-phase supersaturated solid solution with soft magnetic properties. The temperature stability of this sample is investigated in the following.\newline
The sample is exposed to annealing treatments for 150~\degree C, 300~\degree C, 400~\degree C, 600~\degree C and 800~\degree C for 1~h each in a conventional furnace, since previous studies demonstrated major structural changes at these temperatures \cite{bachmaier2015phase, wurster2020microstructural, stuckler2020mfm}. In Fig.~\ref{fig:SEM2}, SEM micrographs of 300~\degree C (a), 600~\degree C (b) and 800~\degree C (c) annealed states are shown. The 300~\degree C annealed state shows a slightly enhanced grain size with respect to the as-deformed state and some dark particles are visible, indicating the presence of a second phase. For the 600~\degree C annealed state, the size of both, the grain size and the second phase particle increases. The 800~\degree C sample shows a completely different microstructure with high phase contrast, indicating a complete demixing of phases, but the grain size is still in the ultrafine-grained regime.
\newline
In Fig.~\ref{fig:SEM2}(d), Vickers hardness is plotted as a function of annealing temperature. An enhanced hardness can be observed for 150~\degree C-, 300~\degree C- and 400~\degree C-annealed states, which is a typical behavior for nanocrystalline materials \cite{volpp1997grain, wang2004effects, renk2015increasing}. The hardness decreases in the 600~\degree C-annealed state due to grain coarsening. Further grain coarsening as well as demixing, i.e. the reduction in solution hardening, lead to an even smaller hardness in the 800~\degree C annealed state.\newline
\begin{figure}
\begin{center}
\includegraphics[width=0.5\linewidth]{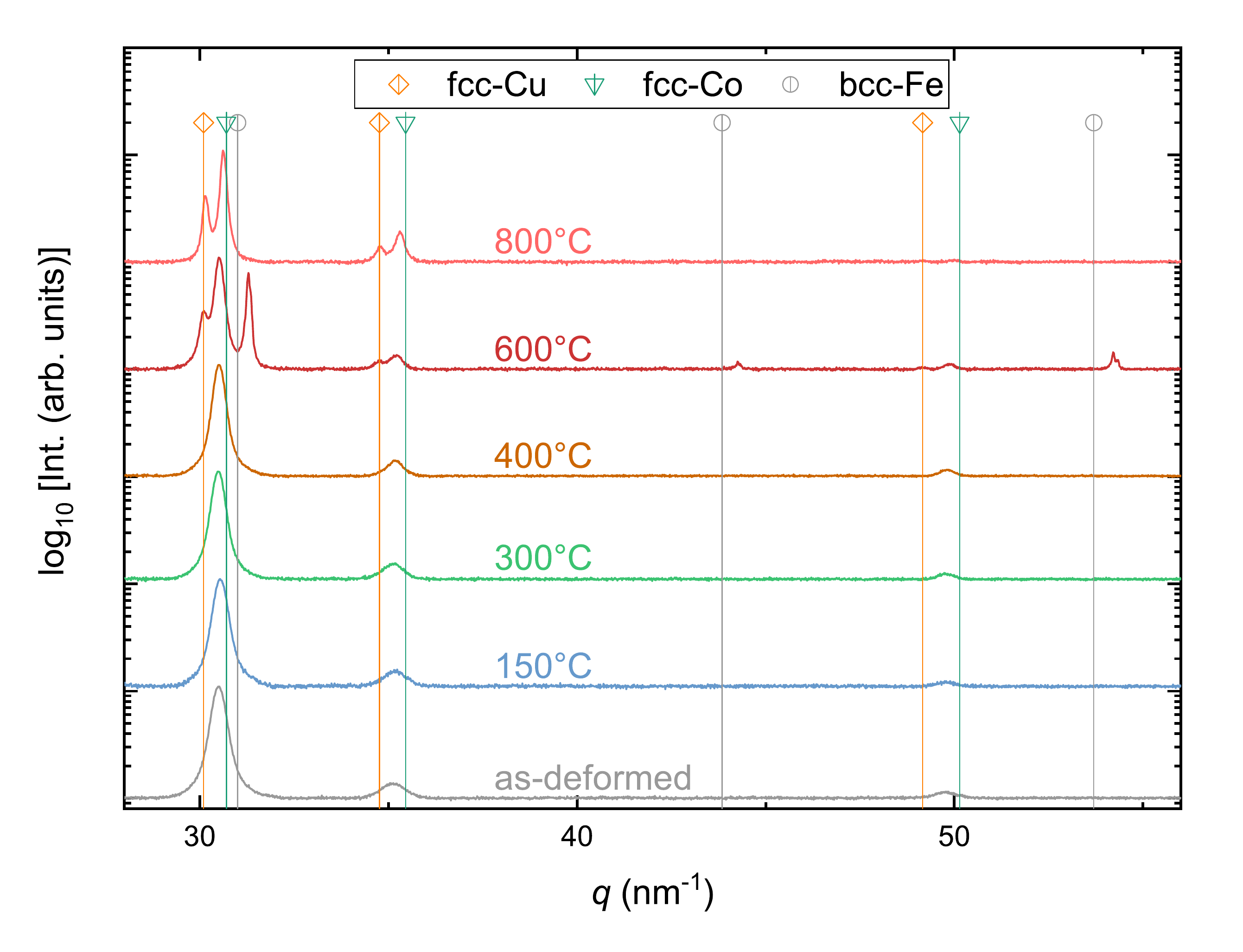}
\end{center}
\caption{XRD-patterns of annealed Cu$_{17}$Fe$_{11}$Co$_{72}$ measured with Co-K$_\alpha$ radiation.}
\label{fig:xrd2}
\end{figure}
To investigate the evolution of crystallographic phases upon annealing in more detail, XRD-measurements were carried out (Fig.~\ref{fig:xrd2}). Starting with the as-deformed state, where only one phase is present, the single-phase fcc-microstructure retains even upon annealing at 400~\degree C. At 600~\degree C, the microstructure decomposes into two fcc-patterns as well as a bcc-pattern forms out. At 800~\degree C, the bcc-pattern vanishes and two fcc-patterns are visible. For a detailed analysis of the formation and decomposition of solid solutions during heat treatment, the lattice constants are evaluated for every pattern by examining the Nelson-Riley function \cite{nelson1945experimental}. Vegard's law \cite{vegard1921konstitution} for solid solutions is extended to oppose different crystal structures by comparing the Wigner-Seitz cell volume, instead of the lattice constant. The Wigner-Seitz cell volume of all appering phases is plotted as a function of annealing temperature $T$ in Fig.~\ref{fig:vegard}. The solid line represents the Wigner-Seitz cell volume for the composition as measured with EDS, coinciding with the value measured with XRD up to 300~\degree C. The Wigner-Seitz cell volume decreases slightly with increasing temperature which is expected to occur due to stress relief. At 600~\degree C, three patterns were observed, whereas one value matches very well with fcc-Cu. We therefore conclude that two solid solutions of (Fe,Co) are present. Assuming the Wigner-Seitz cell volumes of fcc-Co and bcc-Fe, the compositions can be calculated to Fe$_{47}$Co$_{53}$ in the bcc-phase and Fe$_{28}$Co$_{72}$ in the fcc-phase. For annealing at 800~\degree C, the Wigner-Seitz cell volume for Cu remains constant, whereas the bcc-phase vanishes and the value of the second fcc-phase shifts. According to the generalized form of Vegard's law, the composition of the (Fe,Co) solid solution can be calculated to Fe$_{20}$Co$_{80}$. The deviation from the EDS-measurements are expected to arise from any non-linearity in the compositional dependence of the lattice constant. The evolution of crystallographic phases at 600~\degree C and 800~\degree C coincides with the thermal equilibrium Fe-Co phase diagram \cite{ohnuma2002phase, gilles2010probing}, which shows an $\alpha$-(Fe,Co)/$\gamma$-(Fe,Co) two-phase area for 600~\degree C and a single phase $\gamma$-(Fe,Co) solid solution for 800~\degree C for the composition as measured with EDS. Furthermore, a remarkable thermal stability of the supersaturated solid solution up to 400~\degree C is revealed.
\section{Conclusion}
In this study, the processability of ternary Cu-Fe-Co alloys by HPT is investigated starting from initial powder mixtures. Taking results from binary compositions as a basis, which have been successfully processed in recent studies, a region in the ternary phase diagram has been detected, where homogeneous nanocrystalline supersaturated solid solutions can be produced by RT-HPT. For processing of compositions with higher ferromagnetic content, two consecutive steps of HPT-deformation have to be used. The Co-rich specimens form supersaturated solid solution upon HPT-deformation. For the Fe-rich samples, a nanocompound, exhibiting bcc- and fcc-phases, is observed. The nanocompound shows semi-hard magnetic behavior. The formation of $\alpha$-(Fe,Co), exhibiting enhanced magnetic moment, can be observed even in the as-deformed state. For the single-phase supersaturated solid solution, an enhanced magnetic moment cannot be observed but the specimen exhibits soft magnetic behavior. This cannot be explained by grain refinement only, but is expected to arise from a lowering in magnetocrystalline anisotropy. The thermal stability of a specimen with high Co-content shows a persisting nanocrystallinity up to 400~\degree C. At 600~\degree C, decomposition into the stable phases according to thermodynamical equilibrium sets in.
\begin{figure}
\begin{center}
\includegraphics[width=0.5\linewidth]{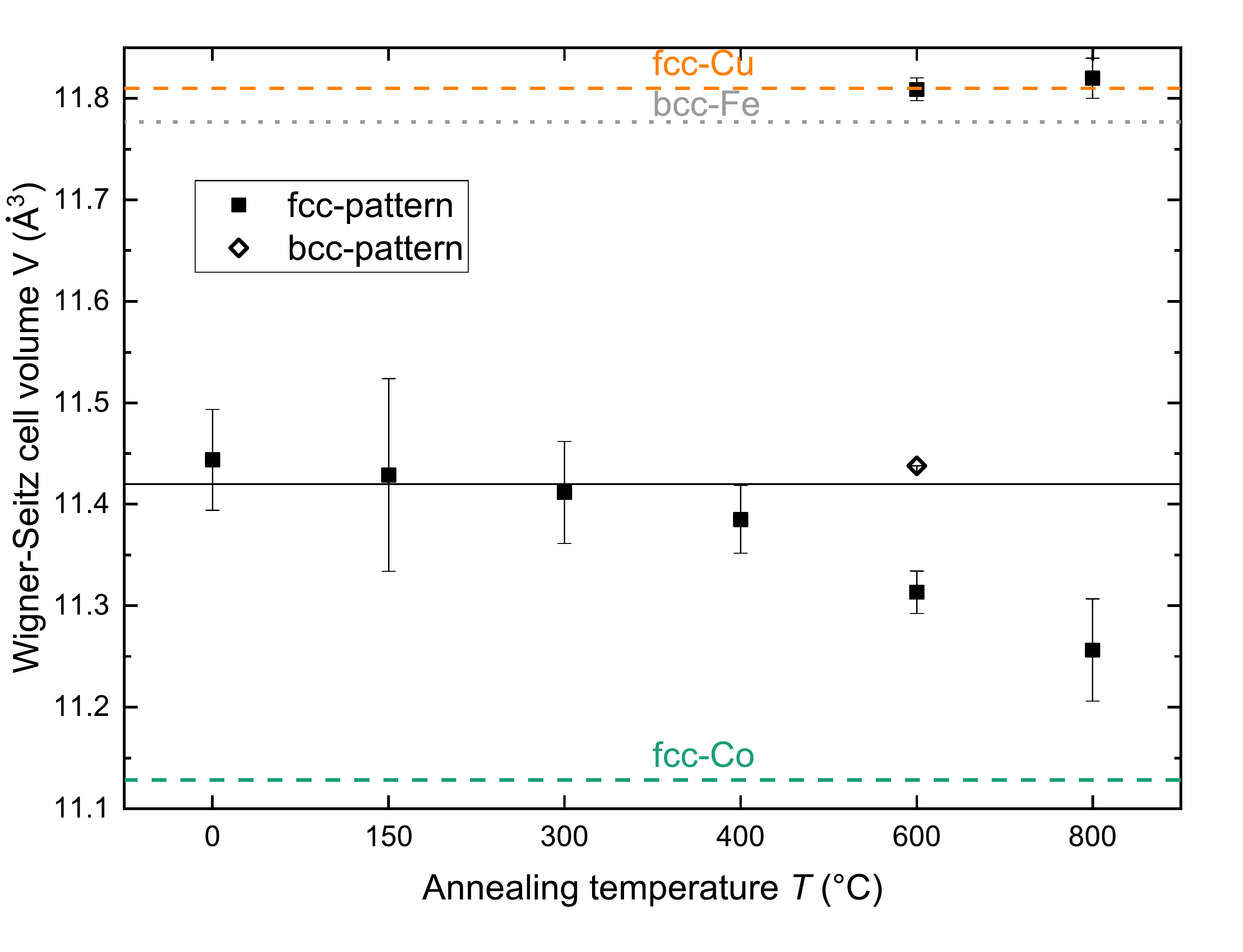}
\end{center}
\caption{Wigner-Seitz cell volume of Cu$_{17}$Fe$_{11}$Co$_{72}$, as determined from XRD-patterns (see Fig.~\ref{fig:xrd2}). The solid line represents the Wigner-Seitz cell volume for the composition measured with EDS.}
\label{fig:vegard}
\end{figure}
\section*{Acknowledgments}
This project has received funding from the European Research Council (ERC) under the European Union’s Horizon 2020 research and innovation programme (Grant No. 757333). The measurements leading to these results have been performed at PETRA III: P07 at DESY Hamburg (Germany), a member of the Helmholtz Association (HGF). We gratefully acknowledge the assistance by Norbert Schell and further thank F. Spieckermann and C. Gammer for their help with data processing. The authors thank M. Kasalo for sample preparation preparation and A. Paulischin for hardness measurements.
\bibliography{libraryjmrt}
\end{document}